\documentclass[a4paper,onecolumn,superscriptaddress,prx, aps, floatfix ]{revtex4-2}
\usepackage{latexsym,amsmath}
\usepackage{amsbsy}
\usepackage[pdftex]{graphicx}
\usepackage{amssymb}
\usepackage{epstopdf}
\usepackage{verbatim}
\usepackage{physics}

\usepackage[usenames]{color}
\usepackage{capt-of} 
\usepackage{float}
\usepackage{ esint }
\usepackage{hyperref}
\usepackage{graphicx}
\usepackage[normalem]{ulem}

\usepackage{xcolor}

\begin{document}

\title{Statistical Mechanics of Random Hyperbolic Graphs within\\ the Fermionic Maximum-Entropy Framework}

\author{M. \'Angeles \surname{Serrano}}
\email{marian.serrano@ub.edu}
\affiliation{Departament de F\'isica de la Mat\`eria Condensada, Universitat de Barcelona, Mart\'i i Franqu\`es 1, E-08028 Barcelona, Spain}
\affiliation{Universitat de Barcelona Institute of Complex Systems (UBICS), Barcelona, Spain}
\affiliation{Instituci\'o Catalana de Recerca i Estudis Avan\c{c}ats (ICREA), Passeig Llu\'is Companys 23, E-08010 Barcelona, Spain}


\begin{abstract}
The intricate relations between elements in natural and human-made systems sustain the complex processes that shape our world, forming multiscale networks of interactions. These networks can be represented as graphs composed of nodes connected by links and, regardless of their domain, they share a set of fundamental structural properties. The family of network models in hyperbolic space constitutes one of the most advanced frameworks accounting for such properties, including sparsity, the small-world property, heterogeneity and hierarchical organization, high clustering, and scale invariance under network renormalization transformations. These geometric models also exhibit other intriguing phenomena, such as an anomalous, temperature-dependent phase transition between a geometric and a non-geometric phase. In simple graph representations, where network links are unweighted, the model can be derived within a statistical-mechanics framework by maximizing the Gibbs entropy of the graph ensemble subject to constraints imposed by observations, with links effectively behaving as fermionic particles. In this topical review, I revisit  these derivations previously scattered across different sources and complement them, in order to properly contextualize and consolidate hyperbolic random graphs within the broad framework of the maximum-entropy principle in the statistical mechanics of complex networks. The approach presented here represents the least-biased prediction of the fundamental set of core network properties and establishes a principled framework for analyzing network structure, offering new perspectives and powerful analytical tools for both theoretical and empirical studies.
\end{abstract}

\maketitle

\section{Introduction}\label{secIntro} 
Complexity is transforming the way in which we understand and predict reality. It revives and places into real scientific practice an old, holistic perspective, in which the world cannot always be understood by reducing phenomena to isolated parts. Instead, complexity highlights that the collective behavior of the entire system emerges from interactions and cannot be predicted from the properties of individual components alone, placing the concept of emergence at its very core. Local interactions in complex systems self-organize without centralized control or design, producing patterns, structures, and functions that exist at the edge of chaos--between order and disorder-- and that are vulnerable yet robust, nested, modular, and exhibit memory effects and adaptation. Thus, complexity marks a shift from viewing real systems as single-scaled, centered on components, and controllable, toward seeing them as multsicale relational systems evolving through feedbacks, co-dependencies, and emergent organization.

Within this paradigm, the study of complex networks has emerged as a powerful framework for understanding the structure and function of real-world complex systems composed of discrete, interacting units. From critical infrastructures, such as the Internet, to the molecular machinery of biological cells, the interconnected neurons of the brain, and the web of social relationships, many seemingly disparate systems share a common underlying architecture: a complex network of many interacting units, typically large and presenting specific structural blueprints. Complex networks provide a natural mathematical and computational language for describing such systems, where nodes can represent components, states, or events (e.g., proteins, individuals, computers, climatic variables), and edges stand for interactions or relationships (e.g., biochemical reactions, friendships, data flows, transition probabilities, or causal influences). 

The networked representation is a convenient abstraction for complex environments that sustain physical processes that happen on top. Unlike Euclidean space --the traditional embedding space in the study of physical systems-- or, when discretized, lattices, which provide a natural definition of length and a hierarchy of length scales, complex networks are, crucially, nonlocal: they lack separability due to the presence of long range interactions. These long range interactions induce short global path lengths in network topology allowing efficient communication or influence across the entire system, but hindering the definition of scalable distances. This lack of distance-based locality, specially when combined with local heterogeneity, poses both conceptual and analytical challenges, but also opens new avenues and insights for uncovering universal organizing principles that transcend specific systems. 

An effective way of restoring spatial locality in complex networks, even in the presence of local heterogeneity, is describing networks within a geometric framework~\cite{boguna2021network}. In the latent hyperbolic geometry framework~\cite{serrano_boguna_2022}, hyperbolic space provides the natural embedding for the simultaneously hierarchical and small-world organization of networks and explains the observed features in their structure by virtue of a family of isomorphic geometric network models where hyperbolic space emerges from the combination of popularity and similarity dimensions~\cite{serrano2008self,krioukov2010hyperbolic}. The models incorporate a universal connectivity law that simultaneously encodes short- and long-range connections and operates at all length scales, with the probability of link formation between nodes depending on their distances in the underlying space. Those distances encompass all attributes contributing to the likelihood of connections, including, but not limited to, explicit distances such as geography in networks like the World Trade Web~\cite{garcia-perez2016hidden}, or 3D Euclidean space in human brain connectomes~\cite{Allard2020}. These models can be reverse-engineered to infer maps of real networks in ultra-low dimensional hyperbolic space~\cite{jankowski2023d}, and they are renormalizable~\cite{garcia-perez2018multiscale} thereby accounting for the scale invariance observed in both the structure\cite{garcia-perez2018multiscale,Zheng2019a,barjuan2025multiscale} and the temporal evolution~\cite{Zheng:2021aa} of many real networks.

The goal of this topical review is to revisit the theoretical foundations of geometric network models, previously presented across disparate works, contextualizing and consolidating them within the broad framework of the maximum-entropy principle in the statistical mechanics of complex networks. Focusing on simple networks with unweighted links, each potential edge between a pair of nodes is a binary occupation variable present/absent, so the ensemble is built under an exclusion-like constraint that forbids multiple occupancy of the same state, leading naturally to links effectively displaying fermionic behaviour with Fermi-Dirac-type probabilities for link formation. It is also important to mention that links are fundamentally unlabeled, and so indistinguishable, just like Fermions are, and so they follow the same statistics. I will revise how the maximum-entropy principle provides a unifying statistical framework for deriving progressively more realistic network models, from the classical random graph model and the configuration model to hyperbolic network models. In this context, essential structural properties--such as the number of nodes and links, the degrees of nodes, and higher-order features like clustering--act as sequentially added constraints that determine the ensemble of admissible networks. Finally, I will outline current challenges and future perspectives for applying geometric approaches to the study of real complex systems. 

\section{Key structural properties of networks}\label{secStruct}
A network is the graph representation of a complex system formed of discrete elements. The elements of the system are represented as nodes, and nodes are connected by links representing interactions (nodes and links are often referred to as vertices and edges in the mathematical and computer science literature). By studying the architecture of network connections --the topology of the network-- one can gain insights into the processes that rule connectivity and how this connectivity is organized into patterns, from the local scale of individual elements to the macroscopic scale of the entire system. 

Here, we focus on the simplest representation, in which the network is unipartite and single-layered --only one category of nodes and one category of links--, with no self-loops (i.e., no link connecting a node to itself), binary, unweighted (links are either present or absent, with no associated intensity), and undirected (reciprocal) links, meaning there is no predefined directionality. The full information about the connectivity structure of such a network representation is encoded in the adjacency matrix $\{a_{ij}\}$, whose entries equal 1 if nodes $i$ and $j$ are connected and 0 otherwise. For undirected networks, the adjacency matrix is symmetric, $a_{ij}=a_{ji}$. Given the practical difficulty of working with very large adjacency matrices in real systems, a set of metrics is typically computed from $\{a_{ij}\}$ to characterize network structure.

At the macroscopic scale, a basic requirement for the system to be functional is that a large fraction of the system is mutually reachable, so that global descriptors are well defined. This is captured by the emergence of a giant connected component (GCC). A connected component in a network is a set of nodes in which every pair is joined by a path, and the GCC in a finite network of $N$ elements is the connected component with the largest cardinality. Formally, determining whether a component is giant requires an analysis of how its size scales with $N$. The size of the GCC must scale linearly with the system size, such that the fraction of nodes in the GCC reaches a nonzero limit \(S\!=\!\lim_{N\to\infty}S_N>0\). Intuitively, the existence of a GCC requires that, when following a randomly chosen link, the node reached has on average more than one excess neighbor (i.e., neighbors other than the one we came from) so that the exploration process keeps branching. In this supercritical regime the number of reachable nodes expands from shell to shell, allowing local neighborhoods to proliferate and coalesce into a macroscopic component, rather than terminating quickly as in a subcritical network where branches typically die out.

Notice that certain network properties are only defined for pairs of nodes that belong to the same connected component. Hence, when working with real networks, it is common practice to extract the GCC and discard smaller components as a preliminary step before performing network analysis. One such property is the distance between nodes, which in networks can be quantified by the shortest-path length, defined as the minimum number of links that must be traversed to go from node $i$ to node $j$, $l_{ij}$. The mean shortest-path distance of a connected network is then computed as $\bar{l}=\frac{1}{N(N-1)}\sum_{i\ne j} l_{ij}$, and the diameter is $l_{max}=\max_{\{i,j\}} l_{ij}$. Typically, real networks display a low value of $\bar{l}$ with a distribution sharply peaked around the average, even for very large networks. This points to the existence of long-range connections, which turn networks a small world in which every element can be reached from any other in a small number of link hops. 

Formally, a network is said to have the small-world property if the number of nodes grows exponentially fast with the diameter of the network, $N \sim e^{l_{max}}$, or, equivalently, the diameter of the network grows slower than any polynomial of the system size $l_{max}\sim \ln N$, which coincides with the expectation if links were created completely at random between pairs of nodes. An ultra-small-world network is one where typical distances grow slower than any polynomial of $\ln N$, for instance as $\sim \ln \ln N$, while the diameter still scales as  $l_{max}\sim \ln N$. This behavior is most commonly associated with power-law degree distributions with $2<\gamma<3$ and infinite variance where hubs create very short routes; for $\gamma>3$, typically networks are instead small-world.

In real networks, it is difficult, or directly impossible, to analyze the scaling with system size, and an indication of the small-world property is the average and distribution of the shortest path lengths. Small mean shortest path length, and hence the small-world property, is one of the fundamental characteristics of complex networks in all domains. For instance, human brain connectomes display  average shortest-path lengths below 3~\cite{barjuan2024optimal}; in the social sciences, where the small-world property is often popularized through the notion of six degrees of separation~\cite{milgram1967small,travers1977experimental,samoylenko2023there}, large-scale online social graphs like Facebook have a mean shortest path length about 4.57, corresponding to 3.57 intermediaries (at least among the 1.59 billion people active on Facebook as of February 2016, https://research.fb.com/blog/2016/02/three-and-a-half-degrees-ofseparation/), highlighting the striking complexity of human social relationships. 

At the local scale, the distinctive feature of nodes in simple graphs is their degree, defined as the number of their connected neighbors. For a node $i$, $k_i=\sum_{j=1,N} a_{ij}$ and, consequently,  the average degree of the network can be calculated as
\begin{equation}
\langle k\rangle=\frac{1}{N}\sum_{i}k_i=\frac{2M}{N},
\end{equation}
where $M$ is the number of network links. The density, giving the fraction of present edges among all possible pairs of nodes, is
\begin{equation}
\rho=\frac{1}{N(N-1)}\sum_{i}k_i=\frac{2M}{N(N-1)}.
\end{equation}
Real networks are usually sparse, meaning that $\rho\ll 1$ or $\langle k\rangle \ll N$. Formally, an ensemble of networks is sparse if $M=O(N)$ as $N\to\infty$ or, analogously, the average degree remains bounded, i.e., $\langle k\rangle=O(1)$.

The collection of degrees defines the degree distribution $P(k)$, which denotes the probability that a randomly chosen node has degree $k$; empirically it is estimated from the histogram of observed degrees. Typically, in many real networks the degree distribution has a power-law shape
\begin{equation}
P(k)\propto k^{-\gamma}
\end{equation}
for large values of $k$, with exponent typically in the range $2<\gamma<3$, which makes the distribution heavy-tailed and implies that most nodes have few links but a few nodes, called hubs, have many. The characteristic exponent $\gamma < 3$ also implies a large variance with the second moment of the degree distribution unbounded, i.e., diverging with the system size for large systems. As a result, the network is referred to as scale-free. In practice, a power-law degree distribution is discernible as a straight line in a log-log plot, but rigorous inference of a power-law should use quantitative methods~\cite{clauset2009power,navas2019universality} rather than visual inspection. Nevertheless, many real world networks display heterogeneous power-law or power-law-like degree distributions.

Broadening the scope to properties that involve nodes' neighbors, the degrees of connected nodes in real networks are typically not independent. A way of quantifying this is through the computation of the average nearest-neighbor degree~\cite{pastor2001dynamical}. The average degree of the neighbors of a node $i$ is
\begin{equation}
k_{nn,i}=
\dfrac{1}{k_i}\displaystyle\sum_{j=1}^N a_{ij}\,k_j, 
\end{equation}
which is usually averaged over the set of nodes in the same degree class $\mathcal{N}(k_i=k)$, $\bar{k}_{nn}(k)=\langle k_{nn,i}\rangle_{ i\in\mathcal{N}(k_i=k)}$. If $\bar{k}_{nn}(k)$ increases with $k$ the network is called \emph{assortative}, meaning that nodes of a certain degree class tend to connect to peers of similar degrees, while if it decreases the network is named \emph{disassortative}, meaning that nodes of low degrees tend to connect to nodes with high degree and vice versa. Typically, social networks tend to be assortative while infrastructure, biological, and economic networks tend to be disassortative. Other ways to measure degree-degree correlations are the global assortativity coefficient $r\in[-1,1]$, which is the Pearson correlation of degrees at the two ends of links~\cite{newman2002assortative}, and the rich-club coefficient $\phi(k_T)$~\cite{colizza2006detecting}, measuring whether nodes with degree $>k_T$ are more interconnected than expected.

Higher order degree correlations involve three nodes, four nodes, and so on. For triads, the tendency to form transitive relations can be measured by the global clustering, $C = 3 \times \#\text{triangles}/\text{ \#triples}$~\cite{newman2001random}. Alternatively, the local clustering coefficient measures the fraction of triangles  involving node $i$, $t_i=\sum_{j<k} a_{ij}\,a_{ik}\,a_{jk}$, as compared to the total possible number given its degree,
\begin{equation}
c_i=\frac{2t_i}{k_i(k_i-1)}\qquad (k_i\ge 2),
\end{equation}
and the average clustering coefficient is $\langle c\rangle=\sum_i c_i/N$~\cite{watts1998collective}. The clustering spectrum represents the average of the clustering coefficient over degree classes, $\bar{c}(k)=\langle c_{i}\rangle_{ i\in\mathcal{N}(k_i=k)}$. Values of the clustering coefficient in real networks are bounded in the interval $[0,1]$ and typically exceed those expected in degree-preserving random graph surrogates, reflecting an enhanced tendency toward triadic closure. Notice that, sometimes, a network is said to be small-world when it simultaneously exhibits high levels of clustering relative to a random graph and short average path lengths comparable to a random graph~\cite{watts1998collective}. Here, I prefer to separate clustering and restrict the small-world property to the later.

For a deeper exploration of network structural features, including mesoscopic organization like community and core-periphery structure, the field of network science offers extensive literature~\cite{dorogovtsev2003evolution,newman2010networks,barabasi2016network,menczer2020first}.

\section{Modelling networks within the maximum entropy framework}\label{sec2}
The nontrivial patterns of connectivity observed in real networks can be captured accurately by network models defining the likelihood of connections between nodes. In general, network models enable researchers to distinguish patterns due to specific ordering principles from random fluctuations under constraints and provide principled explanations for observed regularities. Specifically, Exponential Random Graph (ERG) models are a family of statistical models for network structure that define an exponential probability distribution over graphs to model the likelihood of connections between nodes. The core idea is that the probability of observing a given network depends on selected network statistics, and model dependencies and parameters are chosen so that the expected properties of the resulting graph ensemble match these measurements, i.e., constraints, while the ensemble entropy is maximized. Among all distributions that satisfy the constraints, the maximum-entropy approach spreads probability as uniformly as possible over admissible networks and therefore introduces no additional assumptions beyond the enforced constraints, yielding the least biased description consistent with the data. This framework enables a general statistical-mechanical treatment of networks, drawing on equilibrium methods familiar from the study of ensembles of indistinguishable quantum particles. 

Jaynes formalized the maximum-entropy (MaxEnt) principle --selecting the least-biased distribution consistent with known constraints-- in the 1950s and 1960s~\cite{jaynes1957information,jaynes1957informationII}. He emphasized the natural correspondence between statistical mechanics and information theory, arguing that the Gibbs entropy of statistical mechanics and the Shannon entropy of information theory represent the same underlying concept. This perspective later became an inferential backbone applied to network modelling. ERG models were first proposed in the early 1980s by Holland and Leinhardt~\cite{holland1981}. Here, I follow the statistical mechanics approach to networks developed by Juyong Park and M. E. J. Newman~\cite{park2004statistical}, while noting related contributions~\cite{garlaschelli2008maximum}. 

Given an ensemble of graphs $\mathcal{G}$ and a collection of graph observables ${O_i} , i=1,\ldots, r,$ --possibly measured in empirical observations of some real-world network--, $O_i(G)$ is the value of $O_i$ in graph $G \in \mathcal{G}$. Let $P(G)$ be the normalized probability of graph $G$ in the ensemble $\mathcal{G}$, $\sum_{G} P(G) = 1$. The idea is to choose $P(G)$ so that the expectation value of each graph observable within the ensemble is equal to its observed value $ \langle O_i \rangle = \sum_G P(G)O_i(G)$. This is a vastly underdetermined problem in most cases; the number of degrees of freedom in the definition of the probability distribution $P(G)$ is typically huge compared to the number of constraints imposed by the observations. A standard way to resolve this indeterminacy is to select, among all distributions satisfying the constraints, the one that maximizes the Gibbs entropy $S$ (in statistical mechanics, the macrostate with the highest entropy is the one compatible with the greatest number of specific microstates and is therefore overwhelmingly the most probable and stable macrostate of the system and, thus, defines equilibrium). The problem to solve is finding $P(G)$ by maximizing
\begin{equation}
S = - \sum_G P(G)\ln P(G),
\end{equation}
subject to the known constraints
\begin{equation}
\sum_G P(G)O_i(G) = \langle O_i \rangle.
\end{equation}
One can use the method of Lagrange multipliers that allows to find the maximum or minimum of a function subject to constraints. Introducing Lagrange multipliers $\alpha, \{\theta_i\}$, the maximum entropy given the constraints is achieved for the distribution satisfying
\begin{equation}
\frac{\partial}{\partial P(G)} \left[ 
S + \alpha \left(1 - \sum_{G'} P(G')\right) 
+ \sum_i \theta_i \left( \langle O_i \rangle - \sum_{G'} P(G')O_i(G') \right) 
\right] = 0
\end{equation}
for all graphs $G$. This gives
\begin{equation}
P(G) = \frac{e^{-H(G)}}{Z},
\label{GibbsDist}
\end{equation}
where $H(G)$ is a function that can be interpreted as the graph Hamiltonian,
\begin{equation}
H(G) = \sum_i \theta_i O_i(G),
\end{equation}
and $Z$ is the normalization and can be interpreted as the partition function
\begin{equation}
Z = e^{\alpha+1} = \sum_G e^{-H(G)}.
\end{equation}
Finally, the constraint that the average observable over the ensemble must correspond to some fixed value, $\langle O_i \rangle = \bar{O_i}$, helps to determine the corresponding Lagrange multiplier.
Erd\H{o}sThe Gibbs distribution Eq.(\ref{GibbsDist}) is, therefore, an exponential distribution defining the ERG family. This distribution is the unique unbiased choice: given the imposed constraints, it is the unique distribution that encodes the information contained in those constraints, and crucially, does not introduce any additional information~\cite{shannon1948mathematical,tikochinsky1984consistent,shore2003axiomatic}. Alternatives defined on the basis of generalized entropies might be possible, see~\cite{corominas2024typicality}. 

The thermodynamic ERG variables are
\begin{align}
S& =\left[\ln Z + \sum_i\,\theta_i\langle O_i\rangle\right],\\
  \langle O_i \rangle &= -\left(\frac{\partial \ln Z}{\partial \theta_i}\right)_{\{\theta_{j\ne i} =\text{ct}\}}.   
\end{align}

Notice that the Hamiltonian $H(G)$, and hence the partition function $Z$, are specific to the system under study, but the ERG result is general for any ensemble since the constraints have not been determined yet. Given a particular set of observations constraining network structure, ERG gives the best prediction of the expected value of a graph property $f$ in $\mathcal{G}$, 
\begin{equation}
\langle f \rangle = \sum_{G} P(G)\, f(G). 
\end{equation}

The ERG ensemble is, thus, the least biased ensemble we can construct for a network given a particular set of observations. Depending on which observables are constrained, it gives place to different random network models.

\subsection{The classical random graph}
The simplest possible random graph model is known as the Erd\H{o}s-R\'enyi (ER) model~\cite{erdds1959random,erdos1960evolution}, or classical random graph model. The model was discovered in different variants by mathematicians. The Bernouilli form of the ER model was first introduced by R. Solomonoff and A. Rapoport~\cite{Solomonoff1951} and it was soon analyzed by Paul Erd\H{o}s and Alfr\'ed R\'enyi. They proposed to fix the density of connections in a graph or, equivalently, the average degree, which can be realized by distributing M links uniformly over the $N(N-1)/2$ pairs of nodes in the network giving place to the $\mathcal{G}(N,M)$ ensemble. Self-connections and multiple connections may happen unless a specific restriction to avoid them is added. 

Independently and simultaneously, mathematician Edgar N. Gilbert introduced the canonical form named $\mathcal{G}(N,p)$~\cite{gilbert1959random}, in which each potential edge of a graph in the ensemble is chosen to be included in the graph independently of the other edges, with probability $p$, such that the presence or absence of different edges are mutually independent random variables. This fixes the expected number of links to $\langle L \rangle_{ER} = p N(N-1)/2$.  The $\mathcal{G}(N,p)$ and $\mathcal{G}(N,M)$ models tend to behave very similarly when $N$ is large and they have the same expected average degree. It is usually easier to work with $\mathcal{G}(N,p)$.

The ER model can be thought as the maximum entropy ensemble of graphs with a given expected number of links, results can be found in~\cite{park2004statistical}. Identifying links with particles in traditional statistical mechanics, $\mathcal{G}(N,M)$ would correspond to a microcanonical statistical ensemble of graphs with equal probability and fixed number of links, while $\mathcal{G}(N,p)$ would correspond to a grand-canonical statistical ensamble of graphs where it is not the exact number of links but the expected number of links what is fixed. In the exponential random graph framework, the ER graph Hamiltonian is

\begin{equation}
H(G) = \theta M(G),
\end{equation}
with $M(G)=\sum_{i<j} a_{ij}$, which leads to the partition function
\begin{equation}
Z = \sum_{G} e^{-H(G)} = \sum_{\{ a_{ij} \}} e^{\left( - \theta \sum_{i<j} a_{ij} \right)} 
= \prod_{i<j} \sum_{a_{ij}=0}^1 e^{-\theta a_{ij}}
\end{equation}

\begin{equation}
= \prod_{i<j} (1 + e^{-\theta}) = (1 + e^{-\theta})^{\binom{N}{2}}
\end{equation}
with the following relation between parameter $\theta$ and the probability of connection
\begin{equation}
p = \frac{1}{e^{\theta} + 1}.
\end{equation}
Notice that this is equivalent to a Fermi distribution in the grand-canonical ensemble of traditional statistical mechanics where all the microstates have the same energy and the temperature has no role. The probability of a graph $G$ in this ensemble is
\begin{equation}
P(G) = \frac{e^{-\theta M}}{Z} = \frac{e^{-\theta M}}{(1 + e^{-\theta})^{\binom{N}{2}}} 
= p^{M}(1-p)^{\binom{N}{2} - M}.
\end{equation}
Hence, the least biased ensemble when the only constraint is the expected number of links is that formed by graphs in which each possible link appears with independent probability $p$. The entropy reads
\begin{equation}
S_{ER}
= \binom{N}{2}\,S(p)
= -\binom{N}{2}\Big[p\ln p+(1-p)\ln(1-p)\Big],
\end{equation}
where $S(p)$ is the binary entropy of each potential link. In the limit of sparse networks, $p=\left<k\right>/N$ and the dominant term in the thermodynamic limit is $S_{ER} \approx \left<k\right>N \ln N$, and the entropy per link diverges.

The ER model generates small-world networks, where the average topological distance grows logarithmically with the system size, $d_t \sim log N$~\cite{chung2001diameter}. However, it lacks several key properties of real networks. The ER degree distribution is given by a binomial, 
\begin{equation}
P(k) = \binom{N-1}{k} p^k (1-p)^{\, (N-1-k)}, \quad k = 0, 1, 2, \dots, N-1,
\end{equation}
which in the large N, sparse limit ($p=\langle k \rangle/N$), takes on a Poisson form 
\begin{equation}
P(k) = \frac{\langle k \rangle^k e^{-\langle k \rangle}}{k!}.
\end{equation}
The average nearest neighbors degree function reads
\begin{equation}
k_{nn}(k)=\sum_{k'} k'\,P(k'\mid k)=\sum_{k'} k'\,\frac{k'P(k')}{\langle k\rangle}
=\frac{\langle k^2\rangle}{\langle k\rangle}= p(N-2)+1,
\end{equation}
which is independent of \(k\), and approaches $k_{nn}(k)\approx (N-1)p=\langle k\rangle$
in the thermodynamic sparse limit, implying that the degrees of connected nodes are uncorrelated. Also, the global clustering is 
\begin{equation}
C = \frac{\langle k \rangle}{N-1}
\end{equation} 
and vanishes in the thermodynamic limit. These results are at odds with observed correlations in real networks, which are characterized by heterogeneous degree distributions, assortative or disassortative degree correlations, and high levels of clustering.

\subsection{The generalized random graph}
From the perspective of reproducing the properties of real networks, the Configuration Model (CM)~\cite{bender1978asymptotic,molloy1995critical} was introduced by computer scientists M. Molloy and B. Reed with the goal of modeling networks with a specific degree sequence, including heterogeneous degree distributions. It is widely used in network science to study how structure emerges when only the degree distribution is fixed. As a microcanonical version, hard constraints are imposed on the degree sequence such that every graph in the ensemble has exactly the same degree sequence. This is useful when one needs to reproduce an empirical degree distribution exactly, for instance to avoid fluctuations in the largest degrees. This variant yields maximally random networks with a preassigned degree sequence, which are uncorrelated except for structural unavoidable correlations required by graphicality and to avoid multiple connections. Given a degree sequence, the CM is constructed by assigning each node a number of stubs (half-edges) equal to its degree and then randomly pairing stubs to form edges until none remain. The resulting network matches the prescribed degrees but is otherwise random. In its basic form, it may generate self-loops or multiple edges, which can be accepted or rejected if a simple graph is required.
  
Uniform sampling of simple CM realizations can be difficult, especially for highly heterogeneous degree sequences. A common alternative is degree-preserving random rewiring performing edge swaps: starting from a given network---possibly corresponding to a real system including degree correlations, high clustering, and mesoscopic structural features---repeatedly select two edges and swap their endpoints while rejecting moves that create self-loops or multi-edges; after many swaps (typically $\geq E$), the network is effectively randomized while preserving the degree sequence, up to unavoidable structural correlations.
    
A version of the CM can be derived within the maximum entropy approach~\cite{park2004statistical,bianconi2008entropy,garlaschelli2008maximum}, known as the soft configuration model (SCM). Even if strictly speaking the sparse CM and SCM are not totally equivalent~\cite{anand2009entropy,squartini2015breaking} (they are in the case of dense graphs~\cite{chatterjee2011random}) they are in practical terms. Instead of using the number of links to constraint the graphs as in the ER case, in the SCM the nodes' degrees $k_i$ are constrained with Lagrange multipliers $\alpha_i$. Then, the Hamiltonian is $H(G)=\sum_i \alpha_i k_i$, and the probability of a graph within the ensemble is
\begin{equation}
P(G) = \frac{e^{\sum_i \alpha_i \sum_j a_{ij}}}{Z}
= \prod_{i<j} \frac{e^{(\alpha_i + \alpha_j)a_{ij}}}{1 + e^{(\alpha_i + \alpha_j)}}
\end{equation}
 or, equivalently,
 \begin{equation}
P(G) = \prod_{i<j} p_{ij}^{a_{ij}}(1-p_{ij})^{{1-a_{ij}}} \quad \text{with}  \quad p_{ij}=\frac{1}{ e^{-(\alpha_i + \alpha_j)}+1}.
\end{equation}
Notice that this result bears a similar functional form with the probability of occupancy of a single-particle state with energy \(\epsilon\) in a gas of independent fermions at thermal equilibrium, given by the grand-canonical Fermi-Dirac distribution
\begin{equation}
f(E) = \frac{1}{e^{(\epsilon - \mu) / k_B T} + 1},
\label{fermi}
\end{equation}
where $\epsilon$ is the energy of the specific quantum state, $\mu$ is the chemical potential of the system controlling the number of particles, \textbf{$k_B$} is the Boltzmann constant, and $T$ is the absolute temperature. This similarity led Park and Newman~\cite{park2004statistical} to introduce the analogy between the ensemble of edges in the network and a quantum gas of independent fermions such that pairs of vertices correspond to single-particle states and each single-particle state can be occupied by at most one link, similar to a system obeying the Pauli exclusion principle. 

Redefining $\kappa_i=\mu e^{\alpha_i}$ as a hidden degree variable which stands for the expected degree of a node, where $\mu=\sqrt{\left<\kappa\right>N}$,
 \begin{equation}
p_{ij} =\frac{ 1}{\frac{\mu^2}{\kappa_i \kappa_j}+1} \approx  \frac{\kappa_i \kappa_j}{\langle k \rangle N},
\label{pijSCM}
\end{equation}
which is the connection probability of the CM in its soft version, or soft configuration model, where degrees are not given but fixed on average. While $\mu$ here controls the number of links akin to the homonomous variable in the Fermi-Dirac distribution controlling the number of particles, the analogy does not find an explicit correspondence for the temperature variable. 

The entropy is
\begin{equation}
S_{\mathrm{SCM}}
= \sum_{i<j} S(p_{ij})
= -\sum_{i<j}\Big[p_{ij}\ln p_{ij} + (1-p_{ij})\ln(1-p_{ij})\Big],
\label{entropy}
\end{equation}
where $S(p_{ij})$ is the entropy of each potential link. This expression is valid for any model with pairwise independent interactions. Using the probability of connection Eq.(\ref{pijSCM}), the dominant term in the thermodynamic limit is $S_{SCM} \approx \frac{\left<k\right>}{2} N \ln N$~\cite{anand2014entropy} as for the ER model, and the entropy per link diverges. A further important point is that the microcanonical (hard degrees) ensemble and the canonical (soft degrees) ensemble are not always equivalent in the thermodynamic limit. This means that the microcanonical entropy---given by the logarithm of the number of simple graphs realizing the degree sequence, $\Omega(\{k_i\})$, all with uniform probability---and the canonical entropy $S_{\mathrm{SCM}}$ can differ by an amount that remains extensive, and fluctuations in the canonical ensemble may not vanish. This breaking of ensemble equivalence for degree-constrained graph ensembles is discussed in~\cite{squartini2015breaking}.

For power-law hidden degree distributions with exponent $\gamma < 3$, the expected amount of edges between hubs is larger than one. Mathematicians F. Chung and L. Lu originally introduced the form~\cite{ChungLu2002ConnectedComponents,ChungLu2002AverageDistances}
 \begin{equation}
p_{ij} = \text{min}\left(1, \frac{k_i k_j}{\langle k \rangle N}\right),
\end{equation}
which avoids this problem and is known as inhomogeneous random graph. 

Finally, in the hypersoft configuration model (HSCM)~\cite{hoorn2018sparse}, neither degrees nor even their expected degree values are fixed. It is defined as a mixture of ensembles in every of which the entropy is maximized subject to a fixed sequence of expected degrees, thereby introducing an additional source of randomness. The HSCM can be viewed as a probabilistic mixture of canonical ensembles and is therefore a hypercanonical ensemble, since its latent variables are referred to as hyperparameters in statistics. However, this mixture may not be MaxEnt itself with respect to the distribution of expected degrees. 

In any version of the CM, the generated networks are small world: the average shortest-path distance grows logarithmically with the system size as $\bar{l}\sim \ln N$~\cite{newman2001random}. They can also be sparse when the mean degree remains $O(1)$ as $N$ increases, and --by construction-- can reproduce heterogeneous degree distributions. However, the CM still lacks several key properties of real networks. In terms of two-point degree correlations, it is maximally random given the degree sequence and therefore exhibits no genuine  assortative/disassortative mixing beyond structural correlations, which become more evident for scale-free degree sequences especially when the exponent is close to 2. And the clustering vanishes in the thermodynamic limit. For networks with homogeneous degree distributions it is~\cite{newman2001random} $\bar{c}(k) = C = \frac{\left( \langle k^2 \rangle - \langle k \rangle \right)^2}{N\langle k \rangle^3}$, which decays very fast with system size, while in scale-free networks with $2<\gamma<3$ and natural cut-off it behaves as 
$\bar{c}(k)\sim N^{2-\gamma}\ln N$~\cite{colomer-de-simon2012clustering}, which decays slowly when $\gamma \gtrsim 2$.

\section{Maximum entropy framework for random geometric graphs}\label{sec3}
The ER model captures sparsity and the small-world property due to long-range connections while the CM additionally incorporates heterogeneity in the degree distribution. The question now is what the most fundamental model is that simultaneously accounts for sparsity, the small-world property, heterogeneous degree distributions, and degree correlations in the form of clustering. 

The idea of introducing geometric principles to model the observed high levels of clustering in real networks seems natural. Clustering can be understood as the reflection in the topology of the triangle inequality of network nodes living in a latent geometry in which distances determine the likelihood of connections. However, reconciling the discrete and combinatorial world of graphs and a continuous space description as used in the study of physical and information systems is far from trivial. At first glance, complex networks defy our classical geometric intuition. In small-world networks, traditional notions of closeness and locality break down since any node can be reached from any other in just a few steps traveling links in the topology. 

The answer to reconciling complex networks with a geometric formulation~\cite{boguna2021network} is hyperbolic space. Within a hidden-variable framework, network organization can be described using popularity and similarity coordinates in hyperbolic geometry, which naturally accommodates both the small-world property and the hierarchical character of complex networks~\cite{serrano_boguna_2022}. Nodes that are closer in this space are more likely to connect, while the negative curvature allows high-degree nodes, and especially hubs, to span large portions of the network. Thus, geometry in network science does not necessarily refer to physical space; rather, it denotes an abstract, latent space in which each node has a position and inter-node distances encode the amalgam of attributes that determine connection likelihood. As an additional bonus, the geometric MaxEnt model satisfying sparsity, the small-world property, heterogeneous degree distributions, and high levels of clustering exhibits symmetries that manifest in different forms of scale invariance.

In the next subsections, I review the MaxEnt framework applied to geometric random graphs. First, I introduce the homogeneous case, in which all nodes have the same expected degree, in hyperbolic and asymptotically Euclidean geometries, and then proceed to the generalization to heterogeneous degree distributions in the case of asymptotically Euclidean geometry. Details can be found in~\cite{Boguna:2020fj,vanderkolk2022anomalous}.

\subsection{Geometric Ensembles with Homogeneous Node Attributes.} 
Given a set of nodes with a defined set of internode distances $\{x_{ij}, \forall i,j\}$, we define the total number of links and the total energy as the two global graph observables
\begin{equation}
\begin{aligned}
& M(G)=\sum_{i<j} a_{ij},\\
& E(G)=\sum_{i<j} \varepsilon_{ij} a_{ij}
      = \sum_{i<j} f(x_{ij})\,a_{ij},
\end{aligned}
\end{equation}
where the energy of the link between nodes $i$ and $j$, $\varepsilon_{ij}$, is an increasing function, $f(x_{ij})$, of their distance $x_{ij}$ in the underlying space. The only requirement over the set of distances ${x_{ij}}$ is that they have metric properties. For instance, they can derive from points in a Euclidean space or a hyperbolic space. Apart from the associated coordinates, all nodes are identical. 

We start by considering the MaxEnt ensemble under the constraints
\begin{equation}
\langle M\rangle=\bar M,\qquad \langle E\rangle=\bar E.
\end{equation}
Introducing Lagrange multipliers $\mu$ (fixing $\langle M\rangle$; chemical potential) and $\beta$ (fixing $\langle E\rangle$; inverse temperature), the probability of a graph $G$ in the ensemble is

 \begin{equation}
P(G) = \frac{\prod_{i<j} e^{\beta(\mu-\varepsilon_{ij})a_{ij}}}{Z} = \prod_{i<j} p_{ij}^{a_{ij}}(1-p_{ij})^{{1-a_{ij}}},
\end{equation}
that factorizes in independent link probabilities, taking the Fermi-Dirac form
\begin{equation}
p_{ij}=\frac{1}{e^{\beta(\varepsilon_{ij}-\mu)}+1}.
\label{eq:general0}
\end{equation}
This functional form of the probability, analogous to the probability in Eq.~(\ref{fermi}) of a fermion occupying a quantum state with energy $\varepsilon_{ij}$ at temperature $T=1/\beta$, motivates viewing the model as analogous to a quantum gas of fermions at equilibrium in a thermal bath. Long links (large $x_{ij}$, high $\varepsilon_{ij}$) are, thus, exponentially suppressed, while $\mu$ and $\beta$ tune, respectively, the expected number of links and the average energy. The parameter $\beta$ can be interpreted as a measure of geometric coupling. For large values, connections are predominantly short-ranged, so network topology is tightly constrained by the underlying geometry and nodes tend to connect to their nearest neighbors in the latent space. Conversely, for small values, long-range links become more frequent, distances play a reduced role in shaping connectivity, and nodes can connect broadly across the network; in this regime, the geometric coupling is weak.

Notice that the results above are absolutely general and valid for any underlying metric space. To define further the ensemble, one then must choose this space and the function that relates link energy to distance, $\varepsilon_{ij}=f(x_{ij})$. 

\subsubsection{Asymptotically Euclidean Random Geometric Graphs.}
\label{AERGG}
Let nodes lie in a compact, homogeneous $D$-dimensional Riemannian manifold of volume $N$ without boundary. We assume that its curvature vanishes as $N\to\infty$, so that the space becomes locally Euclidean and approaches $\mathbb{R}^D$ in the thermodynamic limit. Examples include a $D$-sphere or $D$-torus whose size grows with $N$ so as to keep the volume equal to $N$ (growing compact $D$-dimensional hyperbolic manifolds without boundary are also admissible, provided their curvature tends to zero in the thermodynamic limit).

Additional constraints are imposed: networks must be \emph{sparse}, as real networks typically are, meaning that their expected average degree, $\langle k\rangle=2\sum_{i<j} p_{ij}/N$, remains a finite positive constant in the limit $N \to \infty$. Another requirement is that networks in the ensemble are \emph{clustered}, meaning that the probability of connection $p_{ij}$ must be independent of the system size for triangles to persist. The two requirements together imply $\lim_{x \to \infty} f(x)>\frac{D}{\beta} \ln x$, and if this inequality is not satisfied, then clustering vanishes when the network is sparse. In addition, the ensemble is required to be \emph{small-world}, implying that the expected maximum link length, $l_{max} \sim N^{1/(\beta-D)}$, exceeds the space diameter $\sim N^{1/D}$. This happens when $\beta<2D$ and $\lim_{x \to \infty} f(x) \sim \ln x$.

\begin{table*}[t]
\centering
\caption{Phase space of asymptotically Euclidean homogeneous sparse random geometric graphs, defined by the inverse temperature $\beta$. ZC: Zero Clustering. FC: Finite Clustering. SW: Small-World behavior. NSW: Non-Small-World behavior.}
\vspace{0.2cm}
\label{tab:example}
\begin{tabular}{l c c c}
\hline
Domain & Homogeneous degrees &Lagrange Multiplier & Probability of Connection\\
\hline
\multicolumn{4}{l}{$\beta \to 0$ recovers the ER model}\\
\hline
$\beta<D$& ZC, SW&$\mu=\frac{1}{\beta}\ln\left(\hat{\mu}_{ng}\left<k\right>^2\right)$& $\mspace{-30mu}p_{ij}\!=\! \left(1+\left[ x_{ij}^{\beta}/(\hat{\mu}_{ng}\left<k\right>^2)\right]\right)^{-1}$\\
&&$\mspace{-30mu}\hat{\mu}_{ng}\!=\!\frac{N^{\beta\!/\!D\!-\!1}}{\langle k\rangle}\tfrac{(D-\beta)\Gamma(\tfrac{D}{2})}{2\pi^{3D\!/\!2\!-\!\beta}}\!\left[\frac{2\pi^{\frac{D+1}{2}}}{\Gamma\!\left(\tfrac{D+1}{2}\right)}\right]^{1\!-\!\beta\!/\!D}$&\\
\hline
$\beta=D$& \multicolumn{3}{l}{Non-Geometric to Geometric Phase Transition}\\
\hline
$D<\beta<2D$ & FC, SW&$\mu=\frac{1}{D}\ln\left(\hat{\mu}\left<k\right>^2\right)$&$\mspace{-30mu}p_{ij}\!=\!\left(1+\left[ x_{ij}/(\hat{\mu}\left<k\right>^2)^{1/D}\right]^{\beta}\right)^{-1}$\\
&&$\hat{\mu}=\frac{\beta \Gamma(\frac{D}{2})\sin(\frac{D\pi}{\beta})}{2\pi^{1+D/2}\,\langle k\rangle}$&\\
\hline
$\beta=2D$& \multicolumn{3}{c}{Small-world to Non-Small-world Phase Transition}\\
\hline
$\beta>2D$  & FC, NSW &$\mu=\frac{1}{D}\ln\left(\hat{\mu}\left<k\right>^2\right)$&$\mspace{-30mu}p_{ij}\!=\!\left(1+\left[ x_{ij}/(\hat{\mu}\left<k\right>^2)^{1/D}\right]^{\beta}\right)^{-1}$\\
&&$\hat{\mu}=\frac{\beta \Gamma(\frac{D}{2})\sin(\frac{D\pi}{\beta})}{2\pi^{1+D/2}\,\langle k\rangle}$&\\
\hline
\multicolumn{4}{l}{$\beta \to \infty$ recovers the RGG model}\\
\hline
\end{tabular}
\end{table*}

Table I summarizes how the different constraints are fulfilled as a function of the inverse temperature $\beta$ when 
\begin{equation}
\epsilon_{ij}=f(x_{ij})=\ln x_{ij},
\end{equation}
thereby defining the phase space of the ensemble. Table I also includes the functional form of the MaxEnt connection probability under the operating constraints in the limit $N\gg1$ where the space becomes Euclidean and approaches $\mathbb{R}^D$.

The connection probability in Table I for $\beta<D$ has been derived here from results for the heterogeneous case in~\cite{Boguna:2020fj}, reviewed later in subsection~\ref{SMSD}. Parameters $\hat{\mu}$ and $\hat{\mu}_{\mathrm{ng}}$, appearing in the connection probability in the geometric and non-geometric regimes, respectively, control the average degree. Parameter $\beta$ calibrates the coupling of the network topology with the underlying metric space and controls the level of clustering. Notice that $\hat{\mu}_{\mathrm{ng}}$, and therefore the connection probability, depends on the system size for $\beta<D$, so clustering vanishes as $N \to \infty$. In the region $\beta>D$, $\hat{\mu}$ is instead a constant, so the connection probability is independent of the system size and networks in the ensemble have finite clustering. Therefore, $\beta=D$ marks a transition between finite and vanishing clustering. This geometric-to-non-geometric phase transition is topological in nature and related to a reordering of network cycles, and displays anomalous features, such as a diverging entropy and atypical finite-size scaling of the clustering coefficient~\cite{vanderkolk2022anomalous}. Moreover, the slow decay of clustering in the non-geometric phase close to the transition implies that this region is quasi-geometric, and some real networks with relatively high clustering may be better described in this quasi-geometric regime~\cite{van2024random}.

Notice that both short-range and long-range links are encoded in a single connection probability, without the need to model them separately. Long-range connections are affected by temperature, being more probable for higher temperatures, i.e., lower $\beta$ values. For $\beta>2D$, long-range connections that span the system become absent and the network ensemble becomes non-small-world. The limiting case $\beta \to 0$ recovers the ER model, while $\beta \to \infty$ recovers the sharp random geometric graph (RGG) model~\cite{gilbert1961random}, in which nodes are randomly placed in the underlying space and pairs are connected if their distance is below a threshold radius. 

The link energy with the logarithmic dependence on distance is the only function that yields networks that are both small-world and clustered networks, encoding a delicate balance in the prevalence of long-range connections in the network. If $f(x_{ij})$ grows faster, too few long-range connections are created and the small-world property is lost; if it grows more slowly, too many are created and clustering vanishes in the resulting graph. 

\subsubsection{Hyperbolic Random Geometric Graphs.}
\label{HRGG}
Now, consider the homogenous node attribute ensemble in hyperbolic space with constant negative curvature. A hyperbolic space is a complete, simply connected, Riemannian manifold, that is homogeneous and isotropic with negative curvature $K=-\zeta^2$. Geometrically, negative curvature makes the sum of the angles in a triangle less than $\pi$ (180$^\circ$), and the angle deficit increases with the triangle's area. It also causes parallel geodesics to diverge, such that there are infinitely many distinct geodesics through a point not lying on a given geodesic that never intersect the original one (the hyperbolic parallel postulate). Finally, metric balls (the set of all points within distance certain distance) grow rapidly, meaning that the circumference and area of a ball increase exponentially with its radius rather than polynomially in Euclidean space. Negative curvature thus creates a more voluminous space than Euclidean space as you move outward, which is why hyperbolic spaces naturally support hierarchies and tree-like structures, meaning that trees embed with low distortion compared to Euclidean spaces. The network geometry approach has proven that hyperbolic space is also the natural geometry of real complex networks when there is not only hierarchical structure but also the small-world property and high clustering~\cite{boguna2010sustaining,serrano2012uncovering,garcia-perez:2016,allard2018navigable,Zheng2019a,zheng2021scaling}.

To visualize hyperbolic space onto familiar Euclidean space, there are several different equivalent projections, like the upper half-plane model, the Klein disk, and the Poincar\'e disk~\cite{anderson2005hyperbolic}. In the Hyperboloid model, the $(D+1)$-dimensional hyperbolic space is the upper sheet of the two-sheeted hyperboloid in the $(D+2)$-dimensional Minkowski space-time, 
\begin{equation}
\mathbb{H}^{D+1}=\mathcal{H}^{+}=\Big\{x_0\in\mathbb{R}^1\;;\; \vec{x}\in\mathbb{R}^{D+1}:\ x_0^2-\sum_{k=1}^{D+1}x_k^2=1,\ x_0>0\Big\},
\end{equation}
endowed with the Minkowski metric on the hyperboloid. For convenience, the upper sheet of the hyperboloid is projected in an $(D+1)$-dimensional Euclidean space in $(D+1)$-dimensional spherical coordinates, where the radial coordinate of nodes gives the hyperbolic distance from the origin of coordinates and the angular positions are preserved as in the original hyperboloid representation. In this projection, named the native representation, the hyperbolic law of cosines  gives that the hyperbolic distance $d_{\mathbb{H},ij}$ between two points with radial coordinates $r_i$ and $r_j$ and separated by an angular distance $\Delta \theta_{ij}$, computed via 
\begin{equation}
\cosh d_{\mathbb{H},ij}=\cosh \zeta r_i\,\cosh \zeta r_j - \sinh \zeta r_i\,\sinh \zeta r_j\,\cos\Delta\theta_{ij}.
\end{equation}
Here, we consider hyperbolic space of dimension $D+1$ with constant negative curvature
and spread nodes at random within a $D+1$ dimensional hyperbolic ball of radius $R_\mathbb{H}$ with arbitrary radial and angular coordinates. 

If one assumes that the energy of a link is directly proportional to the hyperbolic distance $\varepsilon_{ij}=\frac{\zeta}{2} d_{\mathbb{H},ij}$, the probability of connection becomes
\begin{equation}
p_{ij}=\frac{1}{1+e^{\frac{\zeta}{2}\beta\left(d_{\mathbb{H},ij}-\mu_\mathbb{H} \right)}}.
\label{eq:general}
\end{equation}
This corresponds to the probability of connection in the random hyperbolic graph model, or $\mathbb{H}^{D+1}$ model~\cite{krioukov2010hyperbolic,budel2024random}, where the parameter $\beta$ controls the level of clustering and the chemical potential $\mu_\mathbb{H}$ is a parameter that controls the expected number of links. Therefore, this result proves that $\mathbb{H}^{D+1}$ is a MaxEnt model with tunable average degree and clustering, and the degree distribution can be controlled through the spatial radial distribution of nodes.

\subsection{Asymptotically Euclidean Ensembles with Heterogeneous Node Attributes.}
\label{SMSD}
In this ensemble, instead of fixing the expected average degree as in the homogeneous case, the expected degree of each node \(i\) is fixed to a prescribed value \(\kappa_i\),
\begin{equation}
\langle k_i\rangle \;=\; \Big\langle \sum_{j} a_{ij} \Big\rangle \;=\; \kappa_i, 
\end{equation}
with the corresponding Lagrange multiplier $\alpha_i$. Typically, the values $\kappa_i$, named the hidden degrees, are random and sampled from a given distribution $\rho(\kappa)$. The ensemble is then a mixture of grand-canonical ensembles where the grand-canonical parameters themselves are random, thus referred as hyperparameters, defining thus a hyper-grand-canonical ensemble. As before, there is an additional constraint on the expected value of the total energy, $\langle E\rangle=\bar E$, with Lagrange multiplier $\beta$, which fixes the average link length. These constraints yield the MaxEnt probability of connection 
\begin{equation}
p_{ij}=\frac{1}{e^{\beta\epsilon_{ij}+\alpha_i+\alpha_j}+1}.
\end{equation}
The degree distribution in this ensemble converges to a mixed Poisson distribution whose shape follows the shape of $\rho(\kappa)$~\cite{boguna2003class}. 

Next, as in the homogeneous case in subsection~\ref{AERGG}, we assume that nodes lie in a compact, homogeneous $D$-dimensional Riemannian manifold of volume $N$ without boundary, whose curvature vanishes as $N\to\infty$, so that the space becomes Euclidean and approaches $\mathbb{R}^D$ in the thermodynamic limit. Networks in the ensemble are further required to be \emph{sparse}, \emph{clustered}, and \emph{small-world}. Sparsity imposes that the expected average degree, $\langle k\rangle=2\sum_{i<j} p_{ij}/N$, remains a finite positive constant in the limit $N \to \infty$. Finite clustering requires that the connection probability $p_{ij}$ is independent of the system size, and the qualitative behavior of clustering in this heterogeneous class of models is the same as in the homogeneous case. Sparsity and clustering combined imply $\lim_{x \to \infty} f(x)>\frac{D}{\beta} \ln x$. If this inequality is not satisfied, then clustering vanishes when the network is sparse. The only energy function that yields, in the thermodynamic limit, absence of degree--degree correlations between hidden degrees in these ensembles, so that any remaining correlations are purely structural, is $f (x)= c\ln x$~\cite{Boguna:2020fj}, where $c$ can always be set to $1$ by an appropriate choice of energy units. In addition, the small-world requirement means that links must be able to connect nodes separated by distances of the order of the space diameter, $\sim N^{1/D}$. This is equivalent to demanding that the expected maximum link length, $l_{max}$, is larger than the space diameter, which happens when $\beta<2D$ or $2<\gamma<3$. Thus, networks will be large-worlds whenever $\beta > 2D$ and $\gamma > 3$.

\begin{table*}[t]
\centering
\caption{Phase space of asymptotically Euclidean heterogeneous sparse random geometric graphs, defined by the inverse temperature $\beta$ and the characteristic exponent of the hidden degree distribution $\gamma$. ZC: Zero Clustering. FC: Finite Clustering. USW: Ultra-Small-World behavior. SW: Small-World behavior. NSW: Non-Small-World behavior.}
\vspace{0.2cm}
\label{tab:example}
\begin{tabular}{l c c c c }
\hline
Domain & $2<\gamma<3$ & $\gamma>3$, $\gamma \to \infty$ &Lagrange Multiplier & Probability of Connection\\
\hline
\multicolumn{5}{l}{$\beta \to 0$ recovers the HSCM model}\\
\hline
$\beta<D$&USW, ZC &  SW, ZC &$\alpha=-\left(\ln \kappa + \frac{1}{2}\ln \hat{\mu}_{ng}\right)$ & $\mspace{-30mu}p_{ij}\!= \!\left(1+\left[ x_{ij}^{\beta}/(\hat{\mu}_{ng}\kappa_i \kappa_j)\right]\right)^{-1}$\\
 &  &  &$\mspace{-30mu}\hat{\mu}_{ng}\!=\!\frac{N^{\beta\!/\!D\!-\!1}}{\langle k\rangle}\tfrac{(D-\beta)\Gamma(\tfrac{D}{2})}{2\pi^{3D\!/\!2\!-\!\beta}}\!\left[\frac{2\pi^{\frac{D+1}{2}}}{\Gamma\!\left(\tfrac{D+1}{2}\right)}\right]^{1\!-\!\beta\!/\!D}$&\\
\hline
$\beta=D$& \multicolumn{4}{l}{Non-Geometric to Geometric Phase Transition}\\
\hline
$D<\beta<2D$ &USW, FC & SW, FC &$\alpha=-\frac{\beta}{D}\left(\ln \kappa + \frac{1}{2}\ln \hat{\mu}\right)$&$\mspace{-30mu}p_{ij}\!=\!\left(1+\left[ x_{ij}/(\hat{\mu}\kappa_i \kappa_j)^{1/D}\right]^{\beta}\right)^{-1}$\\
&  &  &$\hat{\mu}=\frac{\beta \Gamma(\frac{D}{2})\sin(\frac{D\pi}{\beta})}{2\pi^{1+D/2}\,\langle k\rangle}$&\\
\hline
$\beta=2D$& &\multicolumn{3}{l}{Small-world to Non-Small-world Phase Transition}\\
\hline
$\beta>2D$  & USW, FC &  NSW, FC &$\alpha=-\frac{\beta}{D}\left(\ln \kappa + \frac{1}{2}\ln \hat{\mu}\right)$&$\mspace{-30mu}p_{ij}\!=\!\left(1+\left[ x_{ij}/(\hat{\mu}\kappa_i \kappa_j)^{1/D}\right]^{\beta}\right)^{-1}$\\
&  &   &$\hat{\mu}=\frac{\beta \Gamma(\frac{D}{2})\sin(\frac{D\pi}{\beta})}{2\pi^{1+D/2}\,\langle k\rangle}$&\\
\hline
\multicolumn{5}{l}{$\beta \to \infty$ recovers the RHG model}\\
\hline
\end{tabular}
\end{table*}

Table II summarizes how the different constraints are fulfilled as a function of the dimension $D$, the inverse temperature $\beta$, and $\gamma$, where $\gamma$ is the characteristic exponent of the Pareto distribution $\rho(\kappa)$ of hidden degrees, thereby defining the phase space of the ensemble. Table II also includes the functional form of the MaxEnt connection probability under the operating constraints in the limit $N\gg1$, in which the space becomes locally Euclidean and approaches $\mathbb{R}^D$.

The expressions for the parameters $\hat{\mu}$ and $\hat{\mu}_{ng}$, appearing in the connection probability in the geometric and non-geometric regimes, respectively, are exactly the same as in the homogeneous case, and the same analysis applies. Here, these parameters are chosen so that the expected degree of a node with hidden degree $\kappa$ equals its hidden degree, $\bar{k}(\kappa) = \kappa$, and thus they control the degree distribution, while $\beta$ controls the level of clustering. Again, $\hat{\mu}_{\mathrm{ng}}$, and therefore the connection probability, depends on the system size for $\beta<D$, so clustering vanishes as $N \to \infty$. In the region $\beta>D$, $\hat{\mu}$ is instead a constant, so the connection probability is independent of the system size and networks in the ensemble have finite clustering. Therefore, $\beta=D$ marks again the topological geometric-to-non-geometric phase transition between finite and vanishing clustering, with anomalous features such as a diverging entropy and atypical finite-size scaling of the clustering coefficient~\cite{vanderkolk2022anomalous}, as well as quasi-geometric behavior close to the transition~\cite{van2024random}.           

Notice that both short-range and long-range links are encoded in a single connection probability, without the need to model them separately. Long-range connections are also affected by a combination of temperature and degree heterogeneity. Lower temperatures (higher $\beta$) imply mostly short-range links, while higher temperatures increase the likelihood of longer link; in addition, smaller values of the degree-distribution exponent $\gamma$ favor the formation of longer-range connections. For $\beta>2D$ and $\gamma>3$, long-range connections that span the system become absent and the network ensemble becomes non-small-world. When $D<\beta<2D$ and $\gamma>2$, or $\beta>2D$ and $2<\gamma<3$, the model is simultaneously small world and clustered. In the limiting cases, $\beta \to 0$ recovers the HSCM model~\cite{hoorn2018sparse}, while $\beta \to \infty$ recovers the sharp random hyperbolic graph  (RHG) version of the $\mathbb{H}^{D+1}$model~\cite{krioukov2010hyperbolic}, in which the connectivity rule is a deterministic step function of the hyperbolic distance between pairs of points. 

If the link energy of the ensemble is redefined as 
\begin{equation}
\varepsilon_{ij}=\ln\!\left(\frac{x_{ij}}{(\kappa_i\kappa_j)^{1/D}}\right), 
\end{equation}
the connection probability takes the standard Fermi-Dirac form Eq.(\ref{fermi}), which again justifies the correspondence of the model with a gas of fermionic particles in a thermal bath at equilibrium.

Finally, expressions for the entropy per link in $D=1$ were derived in~\cite{vanderkolk2022anomalous}: $S/\left<E\right>$ is essentially a constant that depends only on 
$\beta$ in the geometric regime $\beta>1$; at the transition $\beta=1$, the entropy per link shows anomalous, size-dependent scaling, indicating critical behavior with $\frac{S}{\langle E\rangle}$ no longer extensive but growing with system size containing a leading term $\tfrac{1}{2}\ln N$ plus a slower $\tfrac{1}{2}\ln\ln N$ correction (and also depending on $\langle k\rangle$ and on $\gamma$); and in the non-geometric regime, $\frac{S}{\langle E\rangle}$ increases even more strongly with size with a leading $\ln N$ dependence (and it also depends on $\langle k\rangle$ and $\gamma$, and includes additional $\beta$-dependent constant terms), consistent with many more admissible configurations.
Notice that, in the thermodynamic limit, the entropy of the SCM is recovered when $\beta\to 0$, and the same scalings apply to homogeneous degree distributions ($\gamma\to\infty$) and, as expected, equals that of an ER graph when $\beta\to 0$. 

In any dimension $D$, and $\beta>D$, the entropy per link generalizes to
\begin{equation}
\frac{S}{\langle E\rangle}\simeq
\frac{\beta}{D}-\pi\cot\!\left(\dfrac{D\pi}{\beta}\right),
\end{equation}
and remains independent of the system size, depending only on $\beta/D$, diverging as $1/(\beta-D)$ as it approaches the critical transition $\beta=D$.

\section{Random Geometric Graphs and Hyperbolic Space Models}
The foundational model for the MaxEnt spatial network ensembles reviewed in the last section is the $\mathbb{S}^D$ model~\cite{serrano2008self,vanderkolk2022anomalous,van2024random}. This model is the particular case within the hyper-grand-canonical ensemble where the underlying asymptotically Euclidean metric space, referred to as the similarity space, is a $D$-sphere of radius $R = (N/2\pi^{\frac{D+1}{2}}{\Gamma\left(\frac{D+1}{2}\right)})^{1/D}$, $\left\{x\in\mathbb{R}^{D+1}\ \big|\ \lVert x\rVert^{2}=\hat{R}^{\,2}\right\}$. In the thermodynamic limit, $N \to \infty$, the distribution of nodes is given by a Poisson point process with rate $1$ at positions defined by vectors $\boldsymbol{v}$ in $\mathbb{R}^{D+1}$ with $\lVert \boldsymbol{v}\rVert = R$. The hidden degree of a node, known as the popularity variable and drawn from a given distribution $\rho(\kappa)$ ---typically a power law with exponent $\gamma$ but in general any function (or specific sequence)---, constrains its expected degree. 

The phase space of the $\mathbb{S}^D$ model as a function of the relevant parameters $\beta$ and $\gamma$ in the limit of large systems~\cite{Boguna:2020fj} is thus given here in Table I for homogeneous degree distributions and in Table II for heterogeneous degree distributions. The probability of connection can be written in compact form for any region defined by the inverse temperature $\beta$ in any dimension $D$ as
\begin{equation}
\begin{aligned}
&p_{ij}=\left(1+\frac{x_{ij}^{\beta}}{\left(\hat{\mu}(D)\,\kappa_i\kappa_j\right)^{\text{max}(D,\beta)/D}}\right)^{-1}\,, \\ 
&\hat{\mu}(D)=\hat{\mu}_{ng}\Theta(D-\beta)+\hat{\mu}\Theta(\beta-D)\,, 
\label{eq:sdgeneral}
\end{aligned}
\end{equation}
where $\Theta(\cdot)$ is the Heaviside step function. The distance $x_{ij}=R\Delta \theta_{ij}$ is the similarity distance between the two nodes $i$ and $j$. The corresponding angular separation along the circle in the one-dimensional $\mathbb{S}^1$ model is given by $\Delta \theta_{ij}=\left(\pi-\left|\pi-\left|\theta_i-\theta_j\right|\right|\right)$, where $\theta_i$ defines the angular position of node $i$, and in the $D$-dimensional case it generalizes to $\Delta\theta_{ij}=\arccos\!\left(\frac{v_i\cdot v_j}{\lVert v \rVert^2}\right)$. As discussed above, variable $\hat{\mu}(D)$ determines the average degree and is such that the hidden degree of a node coincides with its expected degree. The inverse temperature $\beta$ calibrates the coupling of the network topology with the underlying metric space and controls the level of clustering.

The $\mathbb{S}^D$ model has a quasi-isomorphic purely geometric formulation in hyperbolic geometry, the $\mathbb{H}^{D+1}$ model introduced above in subsection~\ref{HRGG}. The mapping between the $\mathbb{S}^{D}$ and the $\mathbb{H}^{D+1}$ model goes by transforming the hidden degree of a node into a radial coordinate, while the angular positions of nodes are distributed on the D-sphere as in the $\mathbb{S}^{D}$ model. Originally, the isomorphism for $D=1$ in the region $\beta>D=1$ was introduced in~\cite{krioukov2010hyperbolic}. Later it was generalized for any $\beta$, but still in $D=1$~\cite{van2024random}:
\begin{equation}
r = R_\mathbb{H} - \frac{2 \text{max}(1,\beta)}{\beta \zeta} \ln\!\left(\frac{\kappa}{\kappa_0}\right), \qquad R_\mathbb{H} = \frac{2}{\zeta}\left[\ln\left(\frac{N}{\pi}\right)-\frac{\text{max}(1,\beta)}{\beta} \ln\!\left(\hat{\mu}(1) \kappa_0^2\right)\right],
\end{equation}
where the radius of the hyperbolic space $R_\mathbb{H}$ is defined such as $\kappa(R_\mathbb{H}) = \kappa_0$. And still later on, the transformation was extended to any $D$ in the region $\beta>D$~\cite{desy2022dimension}:
\begin{equation}
r = R_\mathbb{H} - \frac{2}{\zeta D} \ln\!\left(\frac{\kappa}{\kappa_0}\right), \qquad R_\mathbb{H} = \frac{2}{\zeta}\ln 2 \left( \frac{\frac{N}{2\pi^{\frac{1+D}{2}}}\Gamma(\frac{1+D}{2})}{ \hat{\mu} \kappa_0^2}\right)^{1/D}.
\end{equation}

The transformation for any $D$ in the region $\beta<D$ is given here for the first time,
\begin{equation}
r = R_\mathbb{H} - \frac{2}{\zeta \beta} \ln\!\left(\frac{\kappa}{\kappa_0}\right), \qquad R_\mathbb{H} = \frac{2}{\zeta \beta}\ln 2^{\beta}\left(\frac{\left(\frac{N}{2\pi^{\frac{1+D}{2}}}\Gamma(\frac{1+D}{2})\right)^{\beta/D}}{ \hat{\mu}_{ng} \kappa_0^2}\right).
\end{equation}
This leads to the following compact expressions for the transformation in arbitrary dimension $D$ and for any region of the phase space defined by the inverse temperature $\beta$
\begin{equation}
r = R_\mathbb{H} - \frac{2 \text{max}(D,\beta)}{\zeta D \beta} \ln\!\left(\frac{\kappa}{\kappa_0}\right), \qquad R_\mathbb{H} = \frac{2}{\zeta}\!\left[\ln\!\!\left(\frac{2^{f(D)}N^{\frac{1}{D}}\Gamma\left(\frac{D+1}{2}\right)^{\frac{1}{D}}}{\pi^{\frac{1+D}{2D}}}\right)\!-\!\frac{\text{max}(D,\beta)}{D \beta} \ln\!\left(\hat{\mu}(D) \kappa_0^2\right)\right],
\end{equation}
where $\hat{\mu}(D)$ is as in Eq.~(\ref{eq:sdgeneral}) and $f(D)=\Theta(D-\beta)(D-1)/D+\Theta(\beta-D)$. Hence, in all cases, nodes with a larger hidden degree have a smaller radial coordinate and so lie closer to the center of the hyperbolic space. Both in the geometric and non-geometric regime and in any dimension, homogeneity in the distribution of hidden degrees will place nodes in a narrow sector of radial values and the model will keep the spherical geometry of the similarity space, which is asymptotically Euclidean in the thermodynamic limit, while heterogeneity in the distribution of hidden degrees will ensure a radial spread of nodes in the hyperbolic formulation, thereby giving rise to an effective hyperbolic geometry. 

Notice that the curvature of the hyperbolic space sets the units of temperature (it is a scaling factor that can be absorbed into the definition of $\beta$), and one is always free to choose its value as long as it is constant and negative. In the geometric phase of the $\mathbb{S}^{D}$ model, $\beta>D$, it is usually taken to be constant, typically $\zeta=1$. In the non-geometric phase, $\beta<D$, a convenient choice is $\zeta=\beta^{-1}$, which allows one to keep the hyperbolic radius constant even when the temperature goes to infinity in the limit $\beta \rightarrow 0$, where the CM is recovered. This makes the regimes more comparable and visualizations easier.

With the transformation in any $D$ and for any $\beta$, the probability of connection of the $\mathbb{S}^{D}$ model becomes 
\begin{equation}
p_{ij}=\frac{1}{1+e^{\frac{\zeta}{2}\beta \left(x_{ij}-R_\mathbb{H}\right)}}\,,
\end{equation}
which is purely geometric only depending on distances $x_{ij}$. Remarkably, this expression recovers the probability of connection of the $\mathbb{H}^{D+1}$ model Eq.~(\ref{eq:general}), with the radius of the hyperbolic $R_\mathbb{H}$ disk playing the role of the chemical potential and the hyperbolic distance $d_\mathbb{H}$ being approximated as
\begin{equation}
x_{ij}=r_i+r_j+\frac{2}{\zeta}\ln\!\left(\frac{\Delta\theta_{ij}}{2}\right).
\end{equation}
This approximation is very accurate for points separated by $\Delta\theta_{ij} \gg 2\sqrt{e^{-2\zeta r_i}+e^{-2\zeta r_j}}$ when $\zeta r_i, \zeta r_j \gg 1$, and $\Delta\theta_{ij}\ll 1$. The fraction of pairs that violate the convergence of the approximated hyperbolic distance to the hyperbolic distance converges to zero in the thermodynamic limit~\cite{serrano_boguna_2022}. Therefore, the two version are, in practice, isomorphic, showing differences in inter-node distances affecting a microscopic fraction of node pairs, which become irrelevant in the thermodynamic limit, and so without practical implications. 

Remarkably, the $\mathbb{S}^{D}$ and $\mathbb{H}^{D+1}$ are equivalent in practical terms but offer different physical interpretations: the $\mathbb{S}^{D}$ model is Newtonian, focusing on forces of attraction between nodes of masses given by the hidden degrees, while $\mathbb{H}^{D+1}$ is Einstenian, defining connections by pure distance. The $\mathbb{S}^{D}$ model is particularly convenient for analytical calculations because the contributions of the popularity and similarity variables to the effective distance are made explicit, while $\mathbb{H}^{D+1}$ provides a geometric map that is useful for visualization, navigation, and geometric analysis and operations. 

These geometric models play for random geometric graphs the same role that the configuration model plays for random graphs: they represent MaxEnt ensembles with pairwise interactions that produce sparse, small-world networks with possible heterogeneous degree distributions, but in contrast have finite clustering (while remaining hidden degree--degree uncorrelated). These models are highly effective in explaining key features of the connectivity of real networks, such as heterogeneous degree distributions~\cite{serrano2008self,krioukov2010hyperbolic,gugelmann2012random}, significant clustering~\cite{serrano2008self,krioukov2010hyperbolic,gugelmann2012random,candellero2016clustering,Fountoulakis2021}, small-world phenomena~\cite{serrano2008self,abdullah2017typical,friedrich2018diameter,muller2019diameter}, percolation characteristics~\cite{serrano2011percolation,fountoulakis2018law}, spectral aspects~\cite{kiwi2018spectral}, and k-cut self-similarity~\cite{serrano2008self,roya2025}. This geometric framework has also been expanded to explain preferential attachment in growing networks~\cite{papadopoulos2012popularity}, weighted networks~\cite{allard2017geometric}, bipartite networks~\cite{serrano2012uncovering,kitsak2017latent}, networks driven by complementarity~\cite{budel2023complementarity}, multilayer networks~\cite{kleineberg2016hidden,Kleineberg2017}, and networks with community structures~\cite{zuev2015emergence,garcia-perez:2018aa,muscoloni2018nonuniform}. 

The $\mathbb{S}^{D}$ model has been extended to directed networks~\cite{allard2024geometric}, which also admit a link-centric statistical mechanics formulation in which directed links are indistinguishable fermions occupying states $i\to j$, with reciprocal interactions modulated by an energy correction~\cite{boguna2025statistical}. This grand-canonical framework yields closed-form expressions for the partition function, link probabilities, entropy, and reciprocity, reveals a topological phase transition at $\beta=D$ as in the undirected case, and recovers and extends the Directed Configuration Model (DCM)~\cite{park2004statistical,kim2012constructing} and the directed $\mathbb{S}^D$ geometric model~\cite{serrano2008self,krioukov2010hyperbolic,allard2024geometric}, clarifying when reciprocity and clustering vanish or remain finite in the thermodynamic limit. 

\section{Discussion and open problems}\label{sec5}
Maximum-entropy (MaxEnt) network modeling, resulting in Exponential Random Graph (ERG) ensembles, builds a probability distribution over graphs with pairwise links that reproduces a chosen set of observed network properties that constrain its structure while otherwise remaining as random as possible. The idea is to distinguish structure that is genuinely implied by those constraints from what could arise by chance: among all graph distributions consistent with the measured features, the MaxEnt one is the least biased because it does not add extra assumptions beyond the enforced constraints. This viewpoint ties network modeling to statistical mechanics, where equilibrium ensembles are characterized by the constraints they satisfy and by an entropy-maximization principle.

Classic random graph models appear as special cases of this framework. If the only constraint is the average degree (or the overall density of links), the MaxEnt solution is the ER ensemble. It captures sparsity and the small-world property but does not reproduce heterogeneous degrees, degree correlations, or finite clustering. Constraining the degree sequence, or the expected values of degrees, yields the CM family, which can match heterogeneous degree distributions and is maximally random given that information, but still lacks finite clustering in the thermodynamic limit and other correlation patterns commonly observed in real networks.

We have revised here how MaxEnt can be extended to geometric ensembles to incorporate clustering in a principled way, by placing nodes in a latent metric space and letting the likelihood of a link depend on distance in that space. Geometry naturally induces clustering through the triangle inequality: if two nodes are both close to a third, they are likely close to each other. A temperature parameter then controls how strongly distances shape connections, interpolating between geometry-dominated, highly clustered networks and more random, long-range-rich networks with vanishing clustering in a non-geometric regime. This geometric MaxEnt approach provides a minimal route to models that can simultaneously account for sparsity, the small-world property, heterogeneous degrees, and non-vanishing clustering.

In particular the $\mathbb{S}^{D}/\mathbb{H}^{D+1}$ model can be framed within the ERG family. Through its mapping to a (hyper-)grand-canonical ensemble of non-interacting, identical fermions in a thermal bath where link energies depend on inter-node distances in a latent space, the model defines a MaxEnt ensemble that can simultaneously generate sparse small-world clustered networks with arbitrary degree distributions using a distance-dependent connection probability that simultaneously encodes both short- and long-range interactions. Distances in the latent space aggregate all attributes driving connectivity, including explicit spatial constraints in systems such as trade networks~\cite{garcia-perez2016hidden} and brain connectomes~\cite{Allard2020}. This perspective makes clear that network geometry is not an add-on but a foundational layer of complexity: it provides a unified explanation for structure bridging topological regularities with measurable notions of similarity and distance encoded in low-dimensional yet highly informative hidden spaces. 

Interestingly, real networks maps can be obtained by embedding the networks in the latent hyperbolic space via embedding techniques such as \textit{D-Mercator}~\cite{garcia-perez2019mercator,jankowski2023d}, which act as ultra-efficient dimensional reduction method since the required dimensions to embed real networks are very low~\cite{Almagro:2021sf}. In this way, the latent hyperbolic geometry restores an effective notion of locality in real networks where topological distance is intrinsically nonlocal due to the small-world property, reintroducing spatial scale in a meaningful way. This geometric locality provided by hyperbolic network maps can be exploited for different purposes and has opened the door to a genuine geometric renormalization (GR) group for complex networks, with associated scale transformations known for $D=1$ and the geometric regime $\beta>1$~\cite{garcia-perez2018multiscale} and non-geometric regime $\beta\le 1$~\cite{Kolk:2024aa} of the $\mathbb{S}^1/\mathbb{H}^{2}$ model. The GR technique produces networks that belong to the same ensemble at each length scale, with parameters flowing predictably across scales. Specifically, GR produces a flow by iteratively coarse-graining nearby nodes within angular sectors of the similarity space into supernodes and linking supernodes whenever any of their constituent nodes were connected, producing a sequence of $\mathcal{O}(\log N)$ layers that progressively select longer-range connections.  At each scale, the layers preserve degree distributions and clustering while inducing regime-dependent flows of the average degree, including critical boundaries separating small-world and non-small-world behavior~\cite{garcia-perez2018multiscale,Kolk:2024aa}. Complementing GR, the Geometric Branching Growth (GBG) model provides a statistical inverse that fine-grains networks by splitting nodes into descendants and expanding the metric space in a way that preserves the empirical connection law and key structural statistics across scales~\cite{Zheng:2021aa}. Combining GR and GBG a sequence of scaled-up and scaled-down self-similar replicas of real networks can be produced useful for finite-size scaling studies and for simulating expensive dynamics on smaller surrogates. Related extensions address weighted networks through geometric renormalization of weights, producing multiscale weighted replicas~\cite{zheng2024geometric}. 

Looking forward, several challenges remain ahead. Real networks are inherently multiscale, with structure and function intertwined across spatial, temporal, and organizational levels, so understanding how local mechanisms generate global behavior, and how global constraints feed back locally, remains an open pursuit. The renormalizability of the $\mathbb{S}^1/\mathbb{H}^{2}$ MaxEnt ensemble implies that their MaxEnt formulation from a single-scale fits into a scale-consistent generative theory with predictive power about how network structure changes (or stays invariant) across resolutions. It means that when the network is renormalized, the resulting scaled network is still described by the same MaxEnt form, just with rescaled parameters and flow of the average degree. That matters because it makes the model a truly MaxEnt multiscale model where the MaxEnt probability of connection is conserved at each scale. One can then describe connectivity at different resolutions with one consistent MaxEnt theory, instead of fitting a new model at every scale. Self-similar real networks then appear as trajectories (or fixed points) of the renormalization flow where the only relevant feature is the average degree of the network, providing strong evidence that the constraints are the right ones capturing the organizing principles that persist across scales. Renormalizability of the MaxEnt model hence implies predictable parameter flows and the possibility to generate scaled-down networks that preserve key statistics, or invert the flow to build scaled-up replicas, without breaking the MaxEnt consistency, implying that the effective description at each scale is again the least biased one compatible with the coarse-grained constraints. While in many physical systems coarse-graining generates additional effective constraints or new effective correlations, observing the preservation of MaxEnt optimality across scales in the $\mathbb{S}^1/\mathbb{H}^{2}$ ensemble is evidence of a particularly strong self-similar structure and that the constraints are close to a sufficient set already encoding the relevant correlations, or the system is in a universality class where coarse-graining does not require adding new terms.

Increasing realism in the description of real networks brings additional layers of information---weights and signs on links, temporal evolution, multiplex structure, and higher-order interactions---which demand new generative models and inference tools that retain interpretability while balancing accuracy against computational and conceptual complexity. Extending maximum-entropy approaches to these settings is a largely open frontier: while binary, pairwise ensembles admit a clean fermionic interpretation, weighted or multi-edge descriptions naturally suggest bosonic analogues, and higher-order interactions call for ensembles defined on hyperedges or simplices rather than on links. Developing such frameworks in a way that remains analytically tractable, scalable, and compatible with geometric embeddings and renormalization ideas is an important direction for future work within the MaxEnt framework.

Finally, in the realm of prediction, a fundamental tension emerges: how much complexity is necessary to make accurate forecasts? Simple MaxEnt models are attractive because they isolate the explanatory power of a small set of constraints and provide transparent null models, but they may fail when neglected complexities are essential for the phenomenon of interest. Conversely, increasingly realistic models can become computationally expensive and harder to interpret, eroding the very advantages that make MaxEnt appealing. A central challenge, therefore, is to identify the minimal set of constraints (and, when appropriate, the minimal latent geometry) that achieves predictive adequacy for a given problem, and to understand how this set changes across scales, data regimes, and application domains. Balancing realism and rigorous tractability is both an art and a challenging science.

\section{Acknowledgments}
I thank the organizers of the school ``Fundamental Problems in Statistical Physics'' FPSP 2025 in Oropa for creating a stimulating environment and for the opportunity to participate. I am also grateful to Mari\'an Bogu\~n\'a for helpful discussions.

I acknowledge support from Grant PID2022-137505NB-C22 funded by MCIN/AEI/10.13039/501100011033 and by ERDF/EU.


%

\end{document}